\documentclass[jap,preprint,groupedaddress,showpacs]{revtex4}

\usepackage[fleqn]{amsmath}
\usepackage{graphicx}
\usepackage{docs}
\usepackage{bm}
\usepackage[colorlinks=true,linkcolor=blue,citecolor=blue,urlcolor=black]{hyperref}%
\expandafter\ifx\csname package@font\endcsname\relax\else
 \expandafter\expandafter
 \expandafter\usepackage
 \expandafter\expandafter
 \expandafter{\csname package@font\endcsname}%
\fi

\newlength{\Figwidth}
\begin{document}


\title{Study of Thermal Properties of Graphene-Based
Structures Using the Force Constant Method}
\author{Hossein Karamitaheri, Neophytos Neophytou, Mahdi Pourfath, and Hans Kosina}
\affiliation{Institute for Microelectronics, Technische Universit\"at Wien, Gu{\ss}hausstra{\ss}e 27--29/E360, A-1040 Wien, Austria\\
  \hspace*{0.2\linewidth} karami@iue.tuwien.ac.at \hspace{0.2\linewidth}}

\date{\today}

\begin{abstract}
The thermal properties of graphene-based materials are theoretically investigated. The fourth-nea-rest neighbor force constant method for phonon properties is used in conjunction with both the Landauer ballistic and the non-equilibrium Green's function techniques for transport. Ballistic phonon transport is investigated for different structures including graphene, graphene antidot lattices, and graphene nanoribbons. We demonstrate that this particular methodology is suitable for robust and efficient investigation of phonon transport in graphene-based devices. This methodology is especially useful for investigations of thermoelectric and heat transport applications.

\end{abstract}

\maketitle

\section{Introduction}
\label{s:Introduction}

Graphene, a recently discovered form of carbon, has received significant attention over the last few years due to its excellent electrical~\cite{Novoselov04,Liang07,Fiori07,Basu08,Leong11}, optical~\cite{Falkovsky08,Liu11,Yuan11}, and thermal properties~\cite{Balandin08,Ghosh08,Hu10,Ong11}. The electrical conductivity of graphene is as high as that of copper~\cite{Chen08} and the ability of graphene to conduct heat is an order of magnitude higher than that of copper~\cite{Balandin08}. In addition, a large scale method to produce graphene sheets has been reported~\cite{KKim09} which sets the stage for graphene usage in large scale applications. The high thermal conductivity of graphene is mostly due to the lattice contribution, whereas the electronic contribution is much weaker~\cite{Balandin08,Hone99}. Due to its high thermal conductivity, graphene can be especially useful for thermal management applications. On the other hand, graphene-based materials such as roughened nanoribbons~\cite{Sevincli10,Aksamija11}, graphene antidots~\cite{Hu09,Karamitaheri11}, and defected graphene lattices~\cite{Xie11,Zhang11b,Haskins11,Jiang11,Hao11} have been demonstrated to have an extremely low thermal conductivity because of the strong sensitivity of phonon transport to disorder and geometrical imperfections in these channels. The thermal conductivity in non-uniform graphene-based materials is shown to be orders of magnitude below the value of pristine graphene. Such materials would be excellent candidates for thermoelectric applications that require very low values of thermal conductivity.

Recent studies on the thermal conductivity of graph-ene nanoribbons have shown that edge roughness can strongly degrade the thermal conductivity. The results indicate that in the presence of edge disorder, phonon transport can be driven into the diffusive, the phonon-glass, and even the localized regimes~\cite{Sevincli10}. Furthermore, vacancies, defects, and isotope doping have dramatic effects on phonon transmission~\cite{Ouyang09b,Hu10}. Antidot meta-materials could also be employed to design the graphene-based phononic-crystal lattices to achieve specific properties. In such structures, thermal properties such as the phonon density-of-states, group velocity, and heat capacity could be engineered to some extent in a controlled manner~\cite{Paul10}. For this goal to be achieved, proper simulation tools and methodologies, accounting for the relevant nanoscale physics are necessary.

In this work, we use the force constant method (FCM) to describe the dynamics of graphene antidot systems, nanoribbons, and nanoribbons with embedded antidots. To calculate the transport properties we use two methods: i) The Landauer approach~\cite{Rego98} for the 2D periodic antidot systems, and ii) the non-equilibrium Green's function (NEGF) technique~\cite{Datta05Book} for the 1D nanoribbon systems. The NEGF method is usually used for electron transport, however, in this work it is extended to the phonon system~\cite{Wang08}. We show that such methodology is suitable for robust and efficient investigation of phonon transport in graphene-based devices. We find that the thermal conductivity is a strong function of the geometrical features of the channels. Even very small size defects could have a large impact on the thermal conductance. The dependence on the size of the antidots, or their positions in the lattice can be used to design their phonon transport properties with control over a large range of thermal conductivity values.

The paper is organized as follows. In Sec.~\ref{s:Ballistic} we present the FCM and evaluate the phononic bandstructure and ballistic lattice thermal conductance of the antidot grap-hene-based structures. In Sec.~\ref{s:Transport} we use the FCM and the NEGF formalism to investigate thermal transport in nanoribbons and nanoribbon antidot channels. Finally, in Sec.~\ref{s:Summary} we conclude.

\section{Phonon Bandstructure and Ballistic Phonon Transport}
\label{s:Ballistic}

In this section we present the geometry of the graphene-based antidot structure investigated. Then, the FCM for graphene-based structures is introduced. Using this method along with the Landauer formalism, we evaluate the phononic bandstructure and the lattice thermal conductance.

\subsection{Graphene and Graphene Antidot Lattices}
\label{s:Structure}

Low dimensional thermoelectric materials have recently attracted significant attention~\cite{Hochbaum08,Boukai08,Venkatasubramanian01} because they provide the possibility of independently controlling their electronic and phononic properties. Thermoelectric materials must simultaneously have a high Seebeck coefficient, a high electrical conductance, and a low thermal conductance~\cite{Nolas01Book}. A large Seebeck coefficient is achieved in semiconductors with large bandgap. Pristine graphene, on the other hand, is a semi-metal or zero-gap semiconductor and has a low Seebeck value~\cite{Seol10}. Several studies have been conducted on methods to open a bandgap in this material. It has been demonstrated that a bandgap can be introduced by appropriate patterning of the graphene sheet~\cite{Han07,Pedersen08}.

\begin{figure}[h]
  \begin{center}
    \includegraphics[width=0.65\linewidth]{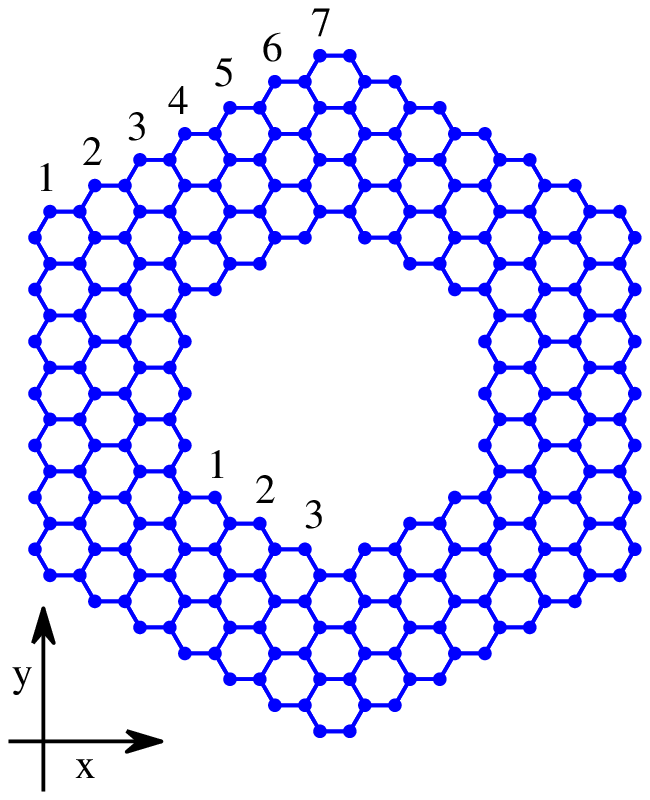}
  \caption{Atomistic geometrical structures of the antidot with $N=7$, and $L=3$, which forms the lattice GAL(7,3).}
  \label{fig:Antidots}
  \end{center}
\end{figure}

One such example of a graphene-based patterned structure is the graphene antidot lattice (GAL) as shown in Fig.~\ref{fig:Antidots}. In this structure, a direct bandgap is obtained depending on the geometrical details of the antidots. In addition, such structure allows for engineering the phonon properties of the material. In the case of thermoelectric materials, the thermal conductivity needs to be drastically reduced. The overall design goal is to identify appropriate geometries that degrade the thermal conductivity and simultaneously improve the power factor, or at least degrade the electronic conductivity much less. In this work we focus on the thermal conductivity part, whereas the electronic part was discussed in our previous work~\cite{Karamitaheri11}.

The unit cell of a GAL can be described by two parameters $L$ and $N$, where $L$ is the side length of the hexagonal unit cell (outer boundary) and $N$ is the side length of the antidot (inner boundary) as shown in Fig.~\ref{fig:Antidots}. Both parameters are described in terms of the graphene lattice constant ($a=2.46\AA$). Figure~\ref{fig:Antidots} shows a hexagonal antidot with $L=7$ and $N=3$, formed by removing $54$ carbon atoms from a cell. This is usually represented using the convention GAL$(7,3)$ as introduced in Ref.~\cite{Pedersen08}.

Here, we consider antidots of hexagonal shape, periodically repeated in the entire 2D plane. In the first step, the dynamic matrix is constructed using the outer hexagonal structure as shown in Fig.~\ref{fig:Antidots}, which is kept fixed. For the antidots of different sizes, we remove the relevant atoms from the cell. For every atom removed, its corresponding column and row are removed from the dynamic matrix. In this work the type of the antidot's boundary edge is zigzag as shown in Fig.~\ref{fig:Antidots}. Other shapes of antidots can have different edge types and circumferential shapes, that can have some effect on the phononic properties as discussed in our previous works~\cite{Karamitaheri11,Karamitaheri11b}. Here, however, we consider only the zigzag edge structures and focus on the computational aspects of the method.

\subsection{Model and Method}
\label{s:Method}

Among all the models used to describe the phonon bands, such as first principle models~\cite{Ye04,Jiang11}, the valence force field (VFF) method~\cite{Lobo97,Kusminskiy09}, and the FCM, the latter has the lowest computation time requirements. In this model, the dynamics of atoms are simply described by a few force springs connecting an atom to its surroundings up to given numbers of neighbors. In contrast, the VFF method is based on the evaluation of the force constants, which requires a much larger computational times. The FCM uses a small set of empirical fitting parameters and can be easily calibrated to experimental measurements. Despite its simplicity, it can provide accurate and transferable results~\cite{Wirtz04,Wang09}. Thus, it is a convenient and robust method to investigate thermal properties of crystals and in particular of graphene nanostructures.

\begin{figure}[h]
  \begin{center}
    \includegraphics[width=0.65\linewidth]{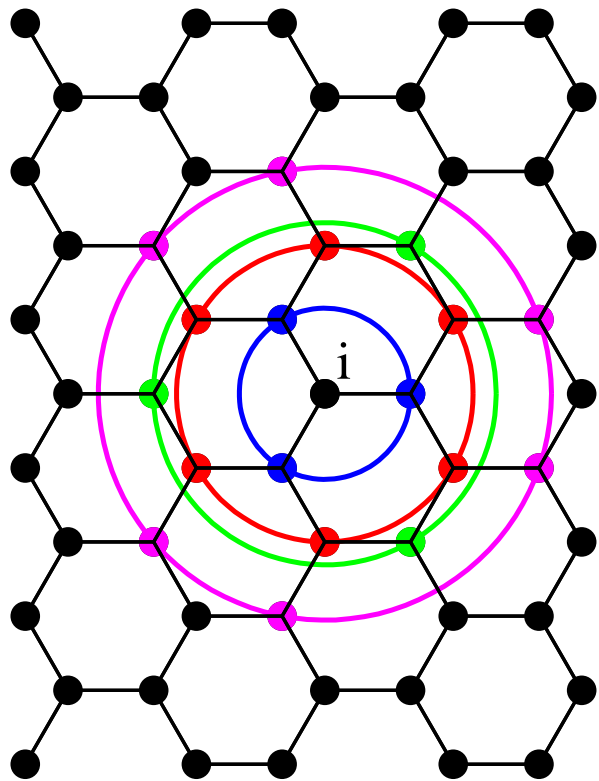}
  \caption{Schematic representation of the nearest neighbors of the $i^{th}$ carbon atom. Up to four nearest-neighbors are included.}
  \label{fig:Nearest}
  \end{center}
\end{figure}

The FCM model we employ involves a fourth nearest-neighbor approximation (see Fig.~\ref{fig:Nearest}). The force constant tensor describing the coupling between the $i^{th}$ and the $j^{th}$ carbon atom, which are the $N^{th}$ nearest-neighbor of each other, is given by:
\begin{equation}
K_0^{(ij)}=\left (\begin{array}{ccc} 
\Phi_r^{(N)} & 0 & 0\\
0 & \Phi_{ti}^{(N)} & 0\\
0 & 0 & \Phi_{to}^{(N)}
\end{array} \right)
\\
\label{e:ForceNN}
\end{equation}
where, $\Phi_r$, $\Phi_{ti}$ and $\Phi_{to}$ are the radial, the in-plane transverse, and the out-of-plane transverse components of the force constant tensor, respectively. Their values are presented in Table~\ref{t:FC}~\cite{Saito98Book}.

The motion of the atoms can be described by a dynamic matrix as: 
\begin{equation}
D=[D_{3\times 3}^{(ij)}]=\left [ \frac{1}{\sqrt{M_iM_j}}\times \left \{ \begin{array}{lll} K_{3\times 3}^{(ij)} & {} & ,i\neq j \\ {} & {} & {} \\-\displaystyle \sum _{l\neq i}K_{3\times 3}^{(il)} & {} & ,i=j \end{array} \right. \right ]
\label{Dyn}
\end{equation}
where $M_i$ is the atomic mass of the $i^{th}$ carbon atom, and $K^{ij}$ is a $3 \times 3$ force constant tensor describing the coupling between the $i^{th}$ and the $j^{th}$ carbon atom. In Cartesian coordinates it is given by:
\begin{equation}
K^{(ij)}=U_m^{-1} K_0^{(ij)} U_m
\label{e:ForceNN2}
\end{equation}
where $U_m$ is a unitary matrix defined as:
\begin{equation}
U_m=\left (\begin{array}{ccc} 
\cos \Theta_{ij} & \sin \Theta_{ij} & 0\\
-\sin \Theta_{ij} & \cos \Theta_{ij} & 0\\
0 & 0 & 1\\
\end{array} \right)
\label{e:Um}
\end{equation}

\begin{table}[tb]
\caption{The fitting parameters of the force constant tensor in $N/\mathrm{m}$ ~\cite{Saito98Book}.}
\label{t:FC}
\begin{center}
\begin{tabular}{|c|c|c|c|}
\hline
\parbox[t]{1.5cm}{\bf N} & \parbox[t]{1.5cm}{\bf $\Phi_r$} & \parbox[t]{1.5cm}{\bf $\Phi_{ti}$} & \parbox[t]{1.5cm}{\bf $\Phi_{to}$}\\
\hline
1 &  365.0 & 245.0  & 98.2\\
2 &  88.0  & -32.3  & -4.0\\
3 &  30.0  & -52.5  &  1.5\\
4 & -19.2  &  22.9  & -5.8\\
\hline
\end{tabular}
\end{center}
\end{table}

Here, we assume that the graphene sheet is located in the $x-y$ plane and that $\Theta_{ij}$ represents the angle between the $x$-axes and the bond between the $i^{th}$ and $j^{th}$ carbon atom. The phononic bandstructure can be calculated by solving the eigen-value problem described by:
\begin{equation}
\left( \sum_{l} K^{(il)}-\omega ^2({\bf k})I \right) \delta _{ij}
- \sum_{l} K^{(il)} \exp{\left ({i{\bf k}\cdot\Delta {\bf r}_{il}}\right )}=0
\end{equation}
where $\Delta{\bf r}_{ij}={\bf r}_i-{\bf r}_j$ is the distance between the $i^{th}$ and the $j^{th}$ carbon atom, and ${\bf k}$ is the wave vector. Equivalently, after setting up the dynamic matrix, one can use the following eigen-value problem:
\begin{equation}
D+\sum_l D_l~\exp{\left({i{\bf k}.\Delta {\bf R}_l}\right )}-\omega ^2({\bf k})I=0
\end{equation}
where $D_l$ is the dynamic matrix representing the interaction between the unit cell and its neighboring unit cells separated by ${\Delta \bf R}_l$.

Using the phononic bandstructures, the density of modes $M(E)$ is calculated, and from this the ballistic transmission $\overline {T}_{\mathrm{ph}}(E)$ is extracted. In the ballistic limit, $\overline {T}_{\mathrm{ph}}(E)$ can be extracted
from the density of modes $M(E)$:
\begin{equation}
\begin{split}
\overline{T}_{\mathrm{ph}}(E)|_{\mathrm{Ballistic}}&=M_{\mathrm{ph}}(E)\\&=\sum_{{\bf k}} \delta (E-\varepsilon_{\mathrm{ph}}({\bf k}))\Delta k_{\perp} \frac{\partial \varepsilon_{\mathrm{ph}}({\bf k})}{\partial k_{\parallel}}
\end{split}
\end{equation}
where $\delta$ is the delta function, $k_{\perp}$ refers to the wave vector component perpendicular to the transport direction and $k_{\parallel}$ to the wave vector component parallel to the transport direction ~\cite{Datta05Book,Kim09}. In our calculations, we broadened the delta function by $1~\mathrm{meV}$. This helps smoothen the numerical results without affecting the results for the thermal conductance. Once the transmission is obtained, the transport coefficient is calculated within the framework of the Landauer theory as~\cite{Ouyang09}:
\begin{equation}
K_\mathrm{ph}=\frac{1}{h}\int_{0}^{+\infty}\overline{T}_\mathrm{ph}(\omega)\hbar\omega\left(\frac{\partial n(\omega)}{\partial T}\right)\ d(\hbar\omega)
\label{e:kp}
\end{equation}
where $n(\omega)$ denotes the Bose-Einstein distribution function. 

\subsection{Computational Results}
\label{s:Results}

\begin{figure}[h]
  \begin{center}
    \includegraphics[width=0.65\linewidth]{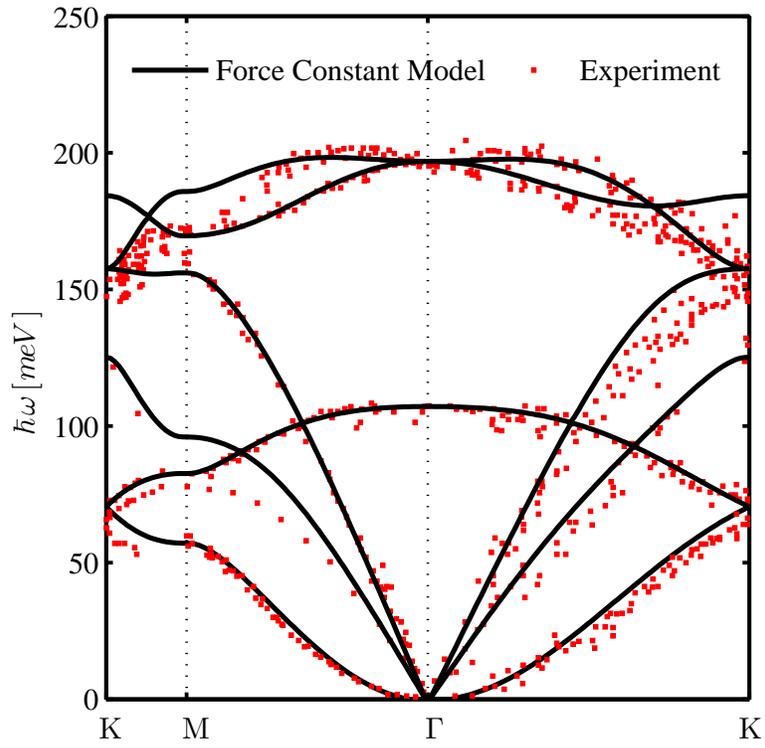}
  \caption{Phononic bandstructure of graphene (solid) evaluated using the fourth
    nearest-neighbor FCM. Experimental results (dots) are
    taken from Refs.~\cite{Wirtz04,Mohr07}.}
  \label{fig:GrapheneBand}
  \end{center}
\end{figure}

The phononic bandstructure of graphene shown in Fig.~\ref{fig:GrapheneBand} is evaluated using the fourth nearest-neighbor FCM with force constants given in Table.~\ref{t:FC}. This method relies on twelve fitting parameters that determine the force constants, which are extracted from experiments. To validate the model, we present the experimental phonon bandstructure results from Refs.~\cite{Wirtz04,Mohr07}. As expected, the result is in good agreement with the experimental data (see Fig.~\ref{fig:GrapheneBand}), especially for the low phonon frequencies, which are the most important ones in determining the thermal conductivity. 

Because the model relies on empirical parameters fitted to experiments, it is much more computationally efficient compared to other atomistic formalisms, such as the valence force field (VFF) method. In the VFF method, for example, the force constants for each atom in the unit cell are calculated, and the simulation time is dominated by dynamic matrix construction~\cite{Paul10}. The approximation in that method comes from the parameters used in the evaluation of the potential energy. For FCM, since force constants are empirical parameters, the construction time of the dynamic matrix is negligible, which makes the computation much more efficient. The simulation time is determined by the solution of the eigen-value problem. The price to pay for improving the accuracy and transferability of the FCM, is that four nearest neighbors need to be included, in contrast to just the next nearest-neighbor in the VFF method. In the graphene lattice this results in 18 neighbors for each atom as shown in Fig.~\ref{fig:Nearest}. In FCM we assume that there is a spring between each carbon atom and its $18$ neighbors. We note that this number is reduced in the case of boundary atoms with less nearest-neighbors. 

\begin{figure*}[h]
  \begin{center}
    \includegraphics[width=0.95\linewidth]{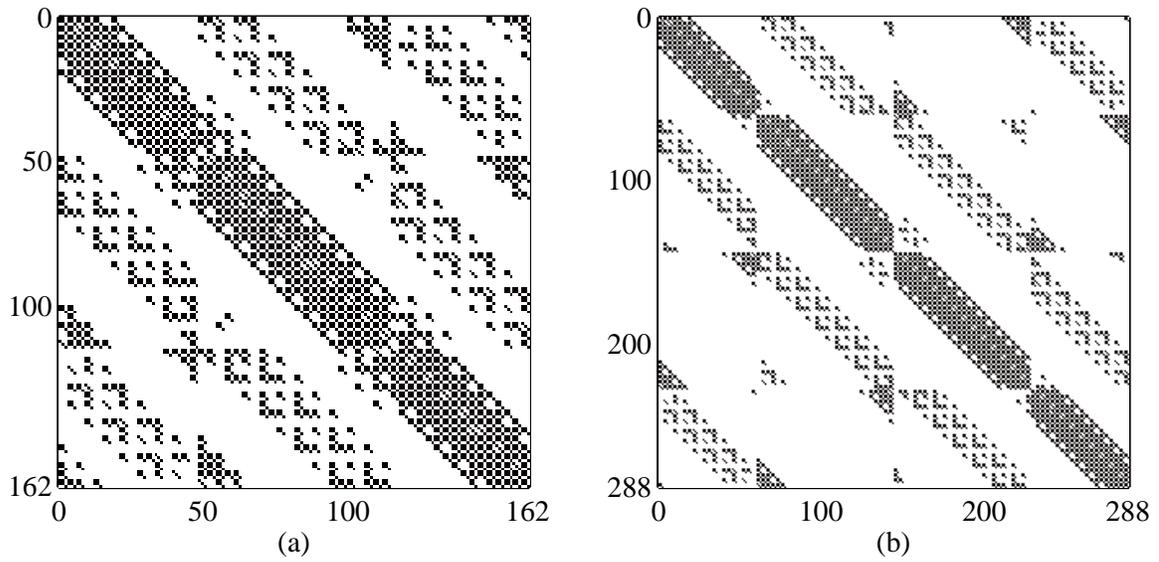}
  \caption{Sparsity pattern of the dynamic matrix of (a) GAL(3,0) and (b)
    GAL(4,0).}
  \label{fig:DM}
  \end{center}
\end{figure*}

In Fig.~\ref{fig:DM} we show two examples of dynamic matrices. Fig.~\ref{fig:DM}-a shows the dynamic matrix for the GAL(3,0) and Fig.~\ref{fig:DM}-b for the GAL(4,0). The sparsity pattern of the dynamic matrix depends on the ordering of the atoms in the physical structure. However, there are specific characteristics associated with the FCM model employed. Since each atom interacts with 18 neighbors, the dynamic matrix has 19 $3\times 3$ blocks filled in each column and row (including the on-site block).

\begin{figure}[h]
  \begin{center}
    \includegraphics[width=0.65\linewidth]{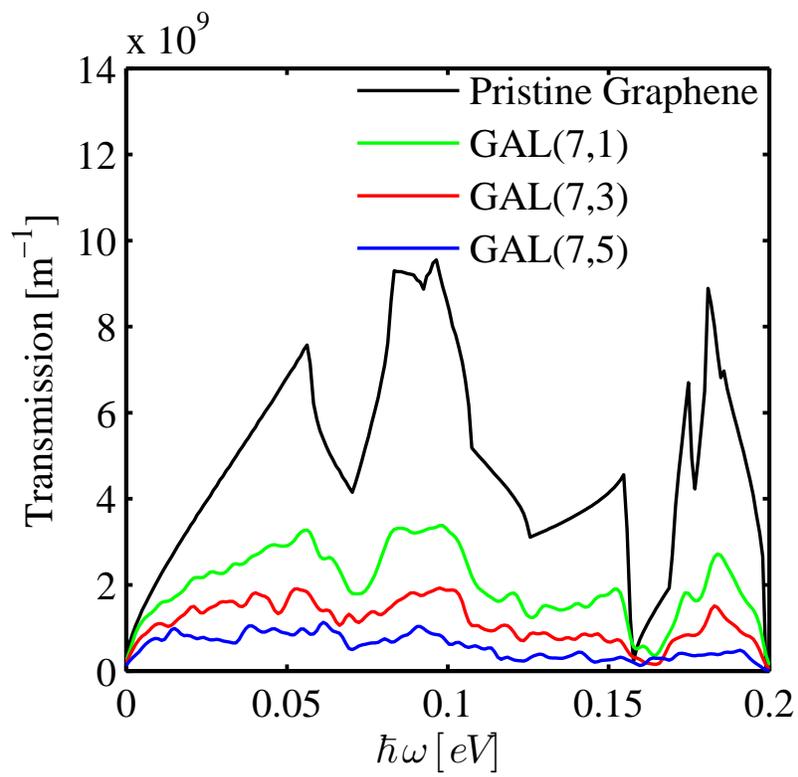}
  \caption{The transmissions of pristine graphene (black)
    and hexagonal GALs of different antidot sizes.}
  \label{fig:HexTrans}
  \end{center}
\end{figure}

Our investigation considers the variation in the thermal conductivity of the GALs upon changes of the geometrical features $L$ and $N$. In Fig.~\ref{fig:HexTrans} we show the phonon transmission of GALs with $L=7$ and different values of $N$. The black-solid line with $N=0$ is the phonon transmission of pristine graphene. The transmission increases almost linearly until $\sim 50~\mathrm{meV}$, where it drops in agreement with other reports in the literature~\cite{Huang10}. This can be easily explained by looking at Fig.~\ref{fig:GrapheneBand}, which shows the lowest phonon mode to extend up to $\sim 50~\mathrm{meV}$ before it reaches the zone boundary. We gradually increase the size of the antidot, and compute the corresponding transmission. As the size of the antidot increases, the phonon transmission is significantly reduced. The total number of atoms in the pristine supercell we consider is 294. In the case of the GAL(7,1) structure, 6 atoms are removed, which is just $2\%$ of the total number of atoms. Even with such a small number of atoms removed, the transmission is reduced considerably as shown by the green line in Fig.~\ref{fig:HexTrans}. As the number of removed atoms increases, i.e. the antidot size increases, the transmission reduces even further. For the GAL(7,3), with 54 atoms removed, which is $\sim 20\%$ of the total number, and for the GAL(7,5) in which 150 atoms are removed ($\sim 50\%$), the transmission monotonically decreases.

The important observation, however, is that the lar-gest degradation in the transmission appears in the first step, where only 6 atoms are removed. After that, the detrimental effect of the antidot size weakens. These results show that the thermal properties of pristine graph-ene are extremely sensitive even to very small geometrical perturbations. 

\begin{figure}[h]
  \begin{center}
    \includegraphics[width=0.65\linewidth]{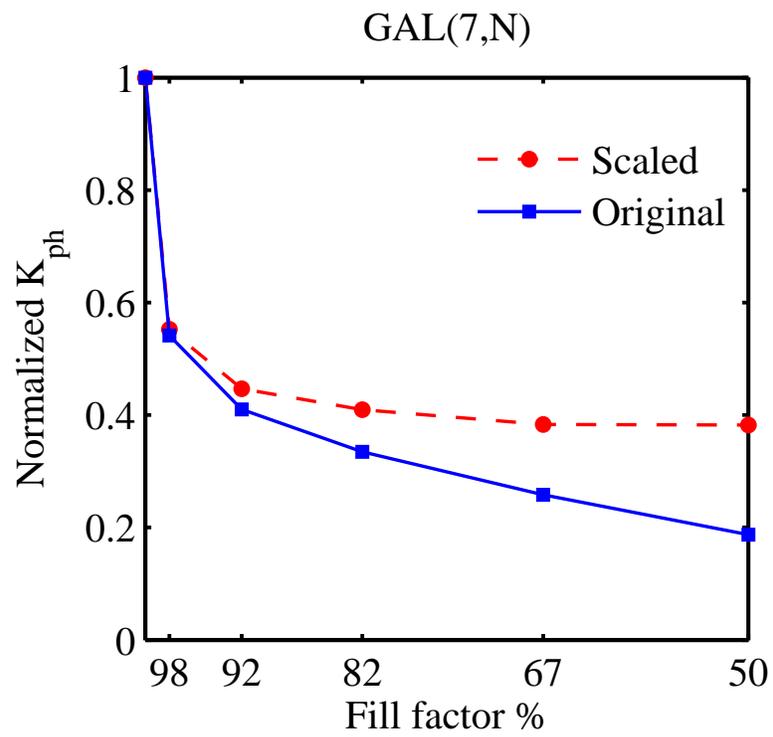}
  \caption{Thermal conductance of GALs of different areas as a function of the antidot filled-factor.}
  \label{fig:TC_Area}
  \end{center}
\end{figure}

When atoms are removed from the lattice, the number of phonon modes could also possibly be reduced. Therefore, a reduction in the transmission would be expected. On the other hand, the drastic reduction in the phonon transmission by small changes in the geometry of the antidots indicates that most of the reduction in the thermal conductance originates from the phononic properties of the lattice, that are changed significantly. The phononic modes are altered, which changes the phonon DOS, their group velocity and possibly introduces strong mode localization as well. To demonstrate this, Fig.~\ref{fig:TC_Area} shows the lattice thermal conductance of different GALs normalized to the thermal conductance of pristine graphene (solid-square line). The result indicates that the conductance is reduced below 60$\%$ as the smallest GAL(7,1) antidots are introduced to form the GAL(7,1). As the size of the antidots increases, the conductance is further reduced, but the rate of decrease weakens. For the GAL(7,5) structure with a fill factor of 50$\%$ the conductance decreases to one-fifth of that of pristine graphene. This strong reduction could have important consequences in the use of such materials for thermoelectric applications, where heat conductivity needs to be minimized.

To illustrate that this effect results from phononic bandstructure engineering, and is not just an effect of a reduced number of modes due to fewer atoms, the dashed-circled line in Fig.~\ref{fig:TC_Area} shows the thermal conductance, but scaled upwards with the filled-factor (FF) as $K_{\mathrm{ph}}/FF$. In this way, we compare the conductance of pristine graphene with that of an antidot lattice with the same number of atoms. The smaller the difference between the two curves, the larger the importance of phononic bandstructure engineering is. It is obvious, that the large reduction in the conductivity introduced by the smaller antidots originates from phononic band modifications. For larger antidots, the reduction of pho-non modes because of the reduced number of atoms might also have some influence.

\section{Coupling FCM to NEGF}
\label{s:Transport}

In this section we couple the FCM and NEGF methods to calculate the thermal conductivity in graphene nanoribbons and graphene nanoribbon antidot channels. Graphene nanoribbons (GNRs) are thin strips of graphene, in which a bandgap forms depending on the chirality of the edges and the width of the ribbon. Electronically, zigzag GNRs (ZGNRs) show metallic behavior, whereas armchair GNRs (AGNRs) are semiconductors with a bandgap inversely proportional to the width. In terms of thermal conductivity, the two configurations show some differences in the order of 30$\%$~\cite{Aksamija11}. The phonon transport properties of nanoribbons have been investigated in the past for pristine~\cite{Xu09,Guo09,Tan11}, rough~\cite{Savin10}, impurity doped~\cite{Jiang10,Jiang11}, or disordered channels~\cite{Haskins11,Xie11}.

Here, we investigate the phonon transport properties in nanoribbons that include antidots. We demonstrate that the FCM can also be effectively coupled to NEGF for the investigation of coherent phonon transport in low dimensional systems. NEGF can be advantageous when it comes to simulating phonon transport in disordered and non-periodic systems. The method has been traditionally employed for electronic transport studies, but has been extended to phonon studies as well~\cite{Zhang07}.

\begin{figure*}[h]
  \begin{center}
    \includegraphics[width=0.95\linewidth]{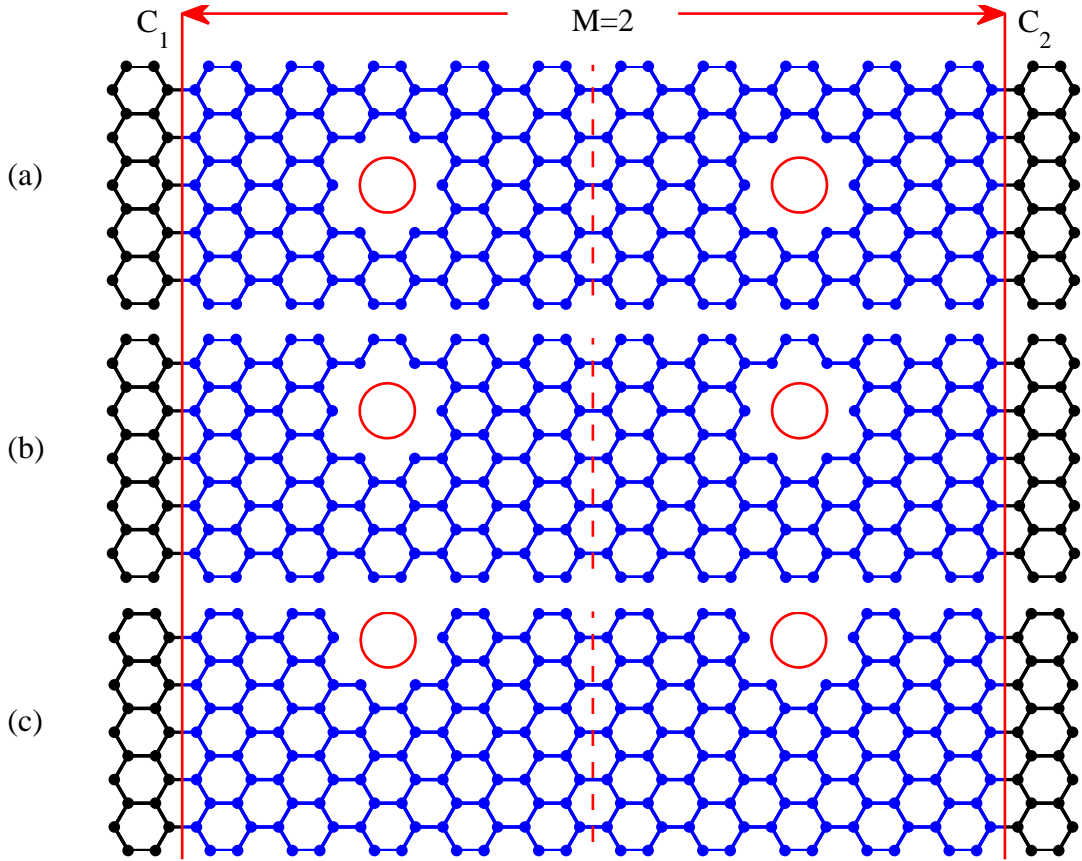}
  \caption{Geometrical structures of nanoribbons with embedded antidots. With respect to the width of the ribbon, the antidots
    are situated at (a) Center, (b) Center+1 atomic layer, and (c) Center+2 atomic layers. $M$ is the
    number of antidots in the channel. $C_1$ and $C_2$ represent the two contacts.}
  \label{fig:GNR}
  \end{center}
\end{figure*}

The system geometry consists of two semi-infinite contacts made of pristine graphene and the device channel including the antidots. The channel length is indicated by $M$, as shown in Fig.~\ref{fig:GNR}, which is determined by the number of antidots placed in the channel. The device is formed by AGNR and the antidots are introduced in the channel part only. The contacts are assumed to be semi-infinite pristine ribbons. In such structure, the calculated thermal properties arise from the channel part of the device only, which breaks the periodicity of the material.    

The device Green's function is obtained by
\begin{equation}
  G(E)=\left (EI-D-\Sigma_{\mathrm{1}}-\Sigma_{\mathrm{1}}
  \right)^{-1}
\end{equation}
where $D$ is device dynamic matrix and $E=\hbar \omega$ is the phonon energy. The contact self-energy matrices $\Sigma_{\mathrm{1,2}}$ are calculated using the Sancho-Rubio iterative scheme~\cite{Sancho85}. The effective transmission probability thro-ugh the channel can be obtained using the relation:
\begin{equation}
  \overline{T}_{\mathrm{ph}}(E)=\mathrm{Trace}[\Gamma_{\mathrm{1}}G\Gamma_{\mathrm{2}}G^{\dagger}]
\end{equation}
where $\Gamma_{\mathrm{1}}$ and $\Gamma_{\mathrm{2}}$ are the broadening functions of the two contacts~\cite{Datta05Book}. The dynamic matrices are constructed using the FCM as explained in Sec.~\ref{s:Method}.

We extract the phonon transmission of the three different structures shown in Fig.~\ref{fig:GNR}. In these structures we place antidots in different positions along the width of the ribbon. For the antidots we consider, we remove 6 atoms, similarly to the GAL(7,1) configuration described in Fig.~\ref{fig:Antidots}. We keep the width of the ribbons constant and introduce antidots in three different places: Center, Center+1 atomic layer, Center+2 atomic layers. Schematics of these structures are shown in Fig.~\ref{fig:GNR}. For each of these structures, we increase the number of antidots ($M$) from one to 10 in a periodic fashion. 

\begin{figure}[h]
  \begin{center}
    \includegraphics[width=0.65\linewidth]{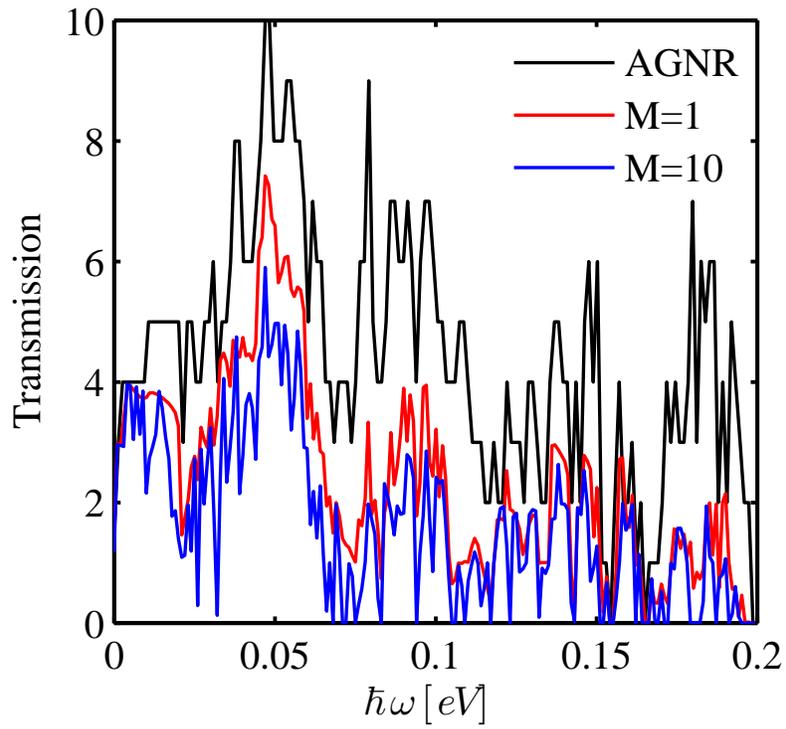}
  \caption{Phonon transmission of ribbons with antidots located at the
    center of ribbons' channel. Black line: No antidots. Red line: One antidot in the channel only. Blue line: 10 antidots in the channel.}
  \label{fig:GNRTrans}
  \end{center}
\end{figure}

The phonon transmission function for the structure in which the antidots are placed in the center of the ribbon is shown in Fig.~\ref{fig:GNRTrans}. The black line shows the transmission function of the pristine ribbon, whereas the red and blue lines are the transmissions of the ribbons that include 1 and 10 antidots, respectively. As with the periodic 2D antidot lattices, by introducing the antidots in the ribbon's channel, the phonon transmission decreases. By introducing antidots the mismatch between the modes in the channel and the contacts increases and thus the transmission and the thermal conductance degrade.

\begin{figure}[h]
  \begin{center}
    \includegraphics[width=0.65\linewidth]{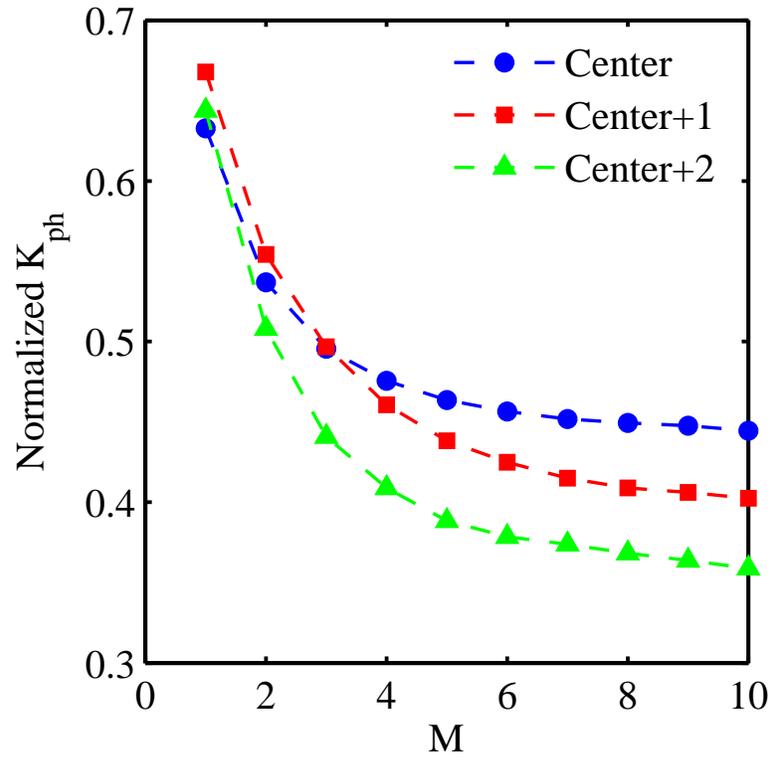}
  \caption{Thermal conductance of ribbons with antidots placed as described in Fig.~\ref{fig:GNR} versus the number of antidots in the channel.}
  \label{fig:GNRTC}
  \end{center}
\end{figure}

In Fig.~\ref{fig:GNRTC} we show the lattice thermal conductance normalized with respect to the pristine ribbon thermal conductance for the three structures in Fig.~\ref{fig:GNR}. The number of antidots in each structure is increased from 1 to 10. Similarly to what was observed in the case of the 2D antidot lattices, the introduction of the first couple of antidots is responsible for most of the thermal conductance degradation. As the number of antidots increases, the rate of degradation decreases. If 10 antidots are introduced the conductance drops to $\sim 40\%$. 

Although for a small number of antidots the conductance is not very sensitive to the antidot placement, as the number of antidots increases, i.e. the channel length increases, some sensitivity of the order of $\pm 10\%$ is observed. The maximum decrease in lattice thermal conductance appears for the case where the antidots are located closer to the ribbon's edge. Other theoretical studies have also concluded that edge defects suppress thermal conductivity significantly~\cite{Savin10}. We note here that randomly placed edge antidots or defects might have a larger degrading effect on thermal conductance than the one observed here for periodic structures since they could drive phonons into localized regimes. Although such studies are not in the focus of this paper, the NEGF technique, applied to phonons using the FCM is perfectly suitable to capture these localization effects. 


\section{Summary}
\label{s:Summary}

We have introduced the fourth nearest-neighbor force constant method to evaluate the phononic properties of graphene antidot lattices. This technique is coupled to the Landauer and the NEGF quantum ballistic transport formalisms. We present the numerical formulation of the method. For the graphene lattice, the ballistic lattice thermal conductance can decrease five times by introducing antidots. Even small size antidots, that reduce the fill factor to only $\sim 98\%$, can have a significant impact. Similar sensitivity to antidots is also observed for nanoribbons. Our results show that the thermal conductivity in armchair graphene-nanoribbons can be significantly reduced in the presence of antidots, which could provide the means for such channels to be efficient thermoelectric materials.

\begin{acknowledgements}
This work was supported by the Austrian Climate and Energy Fund (Contract No. 825467).
\end{acknowledgements}

\bibliographystyle{spphys}       
\bibliography{iue-acronym,IEEEabrv,GrapheneTE}   


\newpage
\clearpage
\vspace*{4cm}

\newpage
\clearpage
\vspace*{4cm}

\newpage
\clearpage
\vspace*{4cm}

\newpage
\clearpage
\vspace*{4cm}

\newpage
\clearpage
\vspace*{4cm}

\newpage
\clearpage
\vspace*{4cm}

\newpage
\clearpage
\vspace*{4cm}

\newpage
\clearpage
\vspace*{4cm}

\newpage
\clearpage
\vspace*{4cm}

\end{document}